\documentclass[%
 reprint, % insert p for preprint
 amsmath,amssymb,
 aps,
pre,
]{revtex4-2}

\usepackage{graphicx}% Include figure files
\usepackage{dcolumn}% Align table columns on decimal point
\usepackage{bm}% bold math
\usepackage{placeins}
\usepackage{color}

\newcommand{\km}{k_\text{max}}
\newcommand{\figabrev}{Fig.}
\newcommand{\multfigabrev}{Figs.}
\newcommand{\eqabrev}{Eq.}
\newcommand{\secabrev}{Sec.}
\newcommand{\refabrev}{Ref.}
\renewcommand{\vec}[1]{\mathbf{#1}}
\usepackage{nicefrac}
\usepackage{hyperref}
\usepackage{tikz}
\usetikzlibrary{math}
\usetikzlibrary{external}
\begin{document}

\title{Neural Density Functional Theory in Higher Dimensions with Convolutional Layers}

\author{Felix Glitsch}
\author{Jens Weimar}
\author{Martin Oettel}
\affiliation{Institute for Applied Physics, University of Tübingen, Auf der Morgenstelle 10, 72076 Tübingen, Germany}

\date{\today}

\begin{abstract}
Based on recent advancements in using machine learning for classical density functional theory for systems with one-dimensional, planar inhomogeneities,
we propose a machine learning model for application in two dimensions (2D) akin to density functionals in weighted density forms, 
as e.~g.\ in fundamental measure theory (FMT).
We implement the model with fast convolutional layers only and apply it to a system of hard disks in fully 2D 
inhomogeneous situations.
The model is trained on a combination of smooth and steplike external potentials in the fluid phase.
Pair correlation functions from test particle geometry show very satisfactory agreement with simulations although these types of external potentials have not been included in the training.
The method should be fully applicable to 3D problems, where the bottleneck at the moment appears to be in obtaining smooth enough 3D histograms as training data from simulations.
\end{abstract}

\maketitle

\tableofcontents

\section{Introduction}
Classical density functional theory (cDFT) offers a very efficient way to calculate structural and thermodynamic properties in equilibrium
for colloidal and molecular systems \cite{bobstart} once the density functional is known.
However, finding such functionals (different for each different type of classical interaction potential between particles) is a tough task. 
Only for hard-body potentials, analytic functionals based on fundamental measure theory (FMT) have been derived that show quantitative agreement with simulations,
see the review by Roth \cite{rothsreview}.
Although there are still recent advances in improving existing analytic functionals for various systems (mostly making use of FMT techniques) 
\cite{mecke2009,wittmann2017,lutsko2020,stopper2018,barthes2024,michaelspaper,melihspaper}, 
a method that can easily be applied to general systems is highly desired.
Machine learning has proven to be a suitable method for this task \cite{sammldft,sammldft2,catsroij,alessandropatchyparticles,alessandropatchyparticles2,dftmlpaircorrmatching,kelley2024,sammuellerpaircorrresponse,sammuellerc1,sammuellerc1howto,sammuellerc1temperature,kampapairpotentials,sammuellerhyperdensity,sammuellerhyperdensityhowto,buiionicfluidsc1,wuanisoc1},
for a recent review see \refabrev~\cite{reviewalessandromartin}.

Among the different approaches, the one by Sammüller et al.\ is very promising. It uses local learning of the map between the 
inhomogeneous density profile and the first-order direct correlation function ($\rho \leftrightarrow c_1$), and exemplary applications to hard rods in 1D \cite{sammuellerc1howto}
and 3D hard spheres with 1D planar inhomogeneities \cite{sammuellerc1} have demonstrated its potential, even surpassing FMT precision 
for hard spheres. The approach has been extended to attractive systems for the fluid phase in the whole temperature-density plane
\cite{sammuellerc1temperature}, learning pair potentials in one dimension \cite{kampapairpotentials} and also to learning
general observables \cite{sammuellerhyperdensity,sammuellerhyperdensityhowto}.
Variants of this approach have also been applied to ionic fluids \cite{buiionicfluidsc1} and anisotropic systems \cite{wuanisoc1}.
The common feature for all works using the local learning scheme is the restriction to inhomogeneities in a planar (flat wall) geometry, i.~e.\
the map $\rho(z) \leftrightarrow c_1(z)$ (where $z$ is the Cartesian coordinate perpendicular to the wall) is learned which does not allow 
the application of the ML functional to a general, inhomogeneous situation.

Here, we address this problem by showing a simple and versatile extension of the local learning technique to 2D for a system of hard disks, where
we construct the corresponding ML network similar to the analytic form of FMT functionals. 
An appealing feature of the proposed network is the use of a structure with convolutional layers only which guarantees a very fast evaluation
using standard ML software \cite{pytorch}. We study hard disks in their fluid phase in order to have a benchmark comparison to a 
FMT functional \cite{martinsfunctional} for which good accuracy for pair correlations \cite{thorneywork2014} and adsorption in hard pores
and around hard tracers \cite{martin2022} has been demonstrated. We do not address the peculiarities of the fluid-solid transition in the hard disk fluid
which involves a first-order fluid-hexatic transition, followed by a continuous transition to a quasi long-range ordered crystal \cite{bernard2011}.

\section{Model Design} 
The local learning introduced by Sammüller et al.\ in \cite{sammuellerc1,sammuellerc1howto} maps a density profile $\rho = \rho(\vec{r})$ 
to the one-body direct correlation function $c_1 = c_1(\vec{r},[\rho])$.
In the framework of DFT, $c_1$ is calculated as the variational derivative of the excess free energy functional $\mathcal{F}_\text{ex}$, 
which from an \textit{ansatz} using weighted densities (as used in FMT) is given by \cite{rothsreview}
\begin{equation}
    \mathcal{F}_\text{ex} [\rho] = k_B \, T \, \int \Phi (\{n_k (\vec{r'})\}) \, \textup{d}^D r',
\end{equation}
where $D$ denotes the dimension. The free energy density $\Phi$ is a nonlinear function of  weighted densities $n_k$, which in turn are convolutions of the density profile $\rho$ and kernels $\omega_k$,
\begin{equation}
    n_k (\vec{r},[\rho]) = (\omega_k \ast \rho) (\vec{r}) = \int \rho(\vec{r'})  \, \omega_k (\vec{r} - \vec{r'}) \, \textup{d}^D r'.
\end{equation}
The quantity of interest is the functional derivative $c_1(\vec{r};[\rho])$, which has the following structure:
\begin{equation}
 \begin{aligned}
     & c_1(\vec{r},[\rho]) = - \beta \, \frac{\delta F_\text{ex}}{\delta \rho (\vec{r})} \\ 
     =& -\!\int \sum_i \left.\frac{\partial \Phi (\{n_k\})}{\partial n_i} \right\vert_{n_k=n_k(\vec{r'},[\rho])} \, \underbrace{\frac{\partial n_i (\vec{r'},[\rho])}{\partial \rho (\vec{r})}}_{\omega_i (\vec{r'} - \vec{r}) =\colon \tilde{\omega}_i (\vec{r}-\vec{r'})} \, \textup{d}^D r' \\
     =& -\sum_i \left(\tilde{\omega}_i \ast \left.\frac{\partial \Phi (\{n_k\})}{\partial n_i} \right\vert_{n_k=\omega_k \ast \rho}\right) (\vec{r}),
 \end{aligned}    
\end{equation}
which with $f_i := -\frac{\partial \Phi}{\partial n_i}$ becomes
\begin{equation} \label{eq:c1eq}
    c_1 (\vec{r},[\rho]) = \sum_i (\tilde{\omega}_i \ast f_i (\{\omega_k \ast \rho\})) (\vec{r}).
\end{equation}

\begin{figure}[h!]
    \centering
    \includegraphics{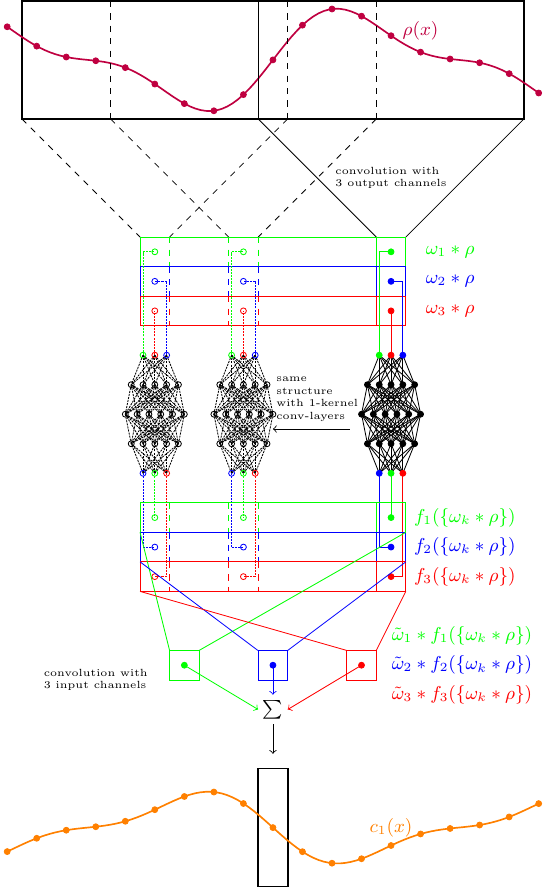}
	\caption{Schematic illustration of the network. For simplicity it is shown for a 1D geometry and only 3 weighted densities ($\km=3$). 
	The figure illustrates the calculation of a single $c_1$-value, i.~e.\ the upper window for $\rho$ is a minimal window size 
	on which the network can be applied.
	Parallel use of the same convolutional kernels is indicated by dashed lines. 
	This includes a whole ``fully connected'' network, which implements the non-linear function $f_i(\{n_k\})$ (see \eqabrev~\eqref{eq:c1eq})
	in the middle between the convolutions with $\omega_k$ and $\tilde{\omega}_k$.}
    \label{fig:networkarchitecture}
\end{figure}

In the ML approach, the mapping between $\rho \leftrightarrow c_1$ is implemented as a convolutional neural network. 
We discuss its architecture by translating the elements of \eqabrev~\eqref{eq:c1eq} into building blocks of a neural network. A schematic visualization for the one-dimensional case with three weighted densities can be found in \figabrev~\ref{fig:networkarchitecture}.
The convolutions $\omega_k \ast \rho$ ($k=1...\km$) are implemented as a convolutional layer with $\km$ channels, acting
on one input channel (density profile). 
The functions $f_i = f_i(\{\omega_k \ast \rho\})$ ($i=1...\km$) are nonlinear real functions of $\km$ real arguments. 
These functions can be realized as a simple multilayer perceptron. However, this perceptron cannot be implemented as a simple stack of 
fully connected layers, as those would expect a fixed input size. Instead, each layer is implemented as a convolutional layer with kernel size 1 
and $\km$ input and output channels, respectively.
This effectively amounts to a parallel execution of a fully connected layer applied to every convolution window from the convolution layer before. 
As a last step, another convolution with the weight functions $\tilde{\omega}_k$ is performed. 
These differ from the weights $\omega_k$ only in the sign applied to the argument $\tilde{\omega}_i (\vec{r}) = \omega_i (-\vec{r})$ 
and are implemented as a similar convolutional layer with $\km$ channels. The weights could be enforced to fulfill $\tilde{\omega}_i (\vec{r}) = \omega_i (-\vec{r})$. 
We opt for not enforcing it (i.e. weights $\tilde{\omega}_k$ are free parameters) such that the FMT form is a possible subset of networks but
there is a higher flexibility for the network to learn the $\rho \leftrightarrow c_1$ map.
Finally, the results of all $\km$ channels are summed together.
As can be seen, the network is constructed using convolutional layers only. Thus, it can be applied to density profiles of arbitrary size, 
independent of the layout of the training data (size of simulation box and size of training windows for the convolutional layers). 
For the range of the convolution (extension of $\omega$ and $\tilde{\omega}$ in each Cartesain direction) we chose $r=\,\sigma$ where $\sigma$ is the diameter of
the hard disks.
Omitting the last convolution is equivalent to the original approach of Sammüller et al. \cite{sammuellerc1howto}, which is discussed later in \secabrev~\ref{sec:sammuellernet}.

In the FMT functional of \refabrev~\cite{martinsfunctional}, we have $\km=7$ for the one-component hard disk fluid, which gives an estimate
of the number of channels needed. However, it should be noted that certain analytic relations connect the different $\omega_k$ in FMT so that the effective
number of independent channels is smaller.
Most results below are generated with $\km=8$, however, the dependency on $\km$ is investigated in \secabrev~\ref{sec:kernels}.

\section{Training the Model}

\subsection{Generating Training Data}
The training data for the neural network are generated by standard Grand Canonical Monte Carlo simulations with insertion, deletion and 
translation moves. 
Similarly to the strategy in Refs.~\cite{sammuellerc1howto,sammuellerpaircorrresponse}, randomly generated external inhomogeneous potentials are 
added, as described in \secabrev~\ref{sec:pots} below. Simulations give access (ground truth data) to both the density profile $\rho(\vec{r})$ (through 2D histograms) and
$c_1(\vec{r})$ through the minimizing equation for the grand potential functional in DFT \cite{bobstart,rothsreview}
\begin{equation} \label{eq:sc}
    c_1 (\vec{r}) = \text{ln} \left(\sigma^2 \, \rho (\vec{r})\right) + \beta \, V^\text{ext} (\vec{r}) - \beta \, \mu,
\end{equation}
with $\mu$ the chemical potential, $\beta = \nicefrac{1}{k_B T}$, $\sigma$ the diameter of the disks and $V^\text{ext}$ the external potential.
The 2D histograms for the density profile $\rho(\vec{r})$ require a higher computational cost than the 1D profiles for e.g. hard spheres, 
since bins are smaller, and for smooth profiles one needs enough counting events in the bins, which leads to longer simulation times.
Therefore, we choose a box length of only 10$\,\sigma$ 
to keep computation effort rather low.
The lateral bin size is $\Delta L'=0.05\,\sigma$, resulting in array sizes $L \times L$ for profiles, with $L=200$.
Some minor finite size effects are visible (as e.g. seen in  $\langle \rho \rangle(\mu)$ for homogeneous systems), 
this however does not affect the performance of the learning scheme and could be remedied by employing larger simulation boxes.

According to \refabrev~\cite{bernard2011}, hard disks show a first-order transition to a hexatic phase at coexisting packing fractions ($\eta = \rho \pi \sigma^2/4$) of
$\eta_\text{fl}=0.700$ and $\eta_\text{hex}=0.716$, corresponding to reduced densities ($\rho^*=\rho\sigma^2$) of $\rho^*_\text{fl}=0.891$ and $\rho^*_\text{hex}=0.912$.
There is a continuous transition to the solid phase at $\eta_\text{cr}=0.720$ ($\rho^*_\text{cr}=0.917$). We have trained the ML model up to average reduced densities in the box
of $\approx 0.75$) which is safely below $\rho^*_\text{fl}=0.891$ in order to avoid interference with the hexatic phase, which is expected to introduce significantly larger 
finite-size errors than those for the fluid phase.
In the remainder, reduced densities are used throughout (i.~e.\ $\rho^\ast \to \rho$).

\subsection{Random Potential Generation}
\label{sec:pots}
Key to good training data is a set of representative inhomogeneous external potentials. In 1D inhomogeneous situations 
randomly generated sine and linear functions as well as hard walls have been used successfully \cite{sammuellerpaircorrresponse}.
In 2D we have an additional ``orientational'' freedom for such 1D potentials.
Fortunately, restricting ourselves to a sum of some rather simple potentials is already sufficient to extrapolate to other geometries outside the training set. 
A generalization of the sine functions used in \cite{sammuellerc1} to 2D is straightforwardly given by
\begin{equation}
	 \label{eq:sine}
    \beta V_\text{sin} (\vec{r}) = \sum_{n=1}^N a_n \, \sin \left(\frac{2 \, \pi}{L'} \, \vec{k}_n \cdot \vec{r} + \phi_n\right),
\end{equation}
with $\vec{k}_n = (k_{x,n},k_{y,n})^T \in \mathbb{N}_0^2$ randomly chosen between 0 and 3, $L'$ the box length, $a_n$ random values for the amplitudes between 0.2 and 1 and the number of sine functions $N$ chosen randomly between 3 and 12.
$\beta\mu$ is chosen between 1 and 7.
In 17 profiles,
hard walls of varying thickness are added at the edges of the simulation box. As the network is trained with local windows only (see following section), this does not make much of a difference.
Implemented in this way, the walls are never curved and are always in a rectilinear geometry.
Additionally, we add locally constant external potentials on a finite domain which we call \textit{plateaus}. 
On the plateau domains a fixed value randomly chosen between 0.2 and 1 is added to or subtracted from $\beta V_\text{sin}$. 
The domain area is generally a freely rotated parallelogram with random side lengths 
between 0.5$\,\sigma$ and 3$\,\sigma$.
However, most profiles with hard walls feature plateaus with an unrotated rectilinear geometry only.
A visualization is shown in the example of \multfigabrev~\ref{fig:beyondthebox} and \ref{fig:plateauproblems} below.
Adding other geometries to $\beta V^\text{ext}(\vec{r})$, especially radial ones, would probably improve the training results, but we opted not to do so in order to use radial geometries for testing the ability to extrapolate to geometries unseen by the neural network.

\subsection{Discretization of the Potential} \label{sec:discretization}
A subtle discretization issue arises when working with discontinuous potentials. The simulations for obtaining training data are in continuous space, and the discrete values of $\rho$ from simulations are averages over the corresponding spatial bins. Thus the simulation maps $\rho \leftrightarrow V^\text{ext}$ and also $c_1 \leftrightarrow V^\text{ext} $  correspond to some suitably averaged external potential over the bin as well. On the other hand, the self-consistent solution of ML-DFT for $\rho$ and $c_1$ from \eqabrev~\eqref{eq:sc} is done on the discrete 2D grid and necessitates a properly discretized $V^\text{ext}$ (corresponding to the above). 
For smooth external potentials like $V_\text{sin}$, one obtains reasonable results by just using the value of the potential at the midpoint of the bin, while for discontinuous potentials (e.g. of plateau-type), this is a problem, when the mid-point of the bin happens to be on the plateau but the plateau covers only part of the bin such that an averaged potential will
be different.
In 1D, one can usually choose the grid and discretization in a way that this problem does not emerge, but in higher dimensions one always has to deal with such problems if one considers arbitrary geometries for discontinuous potentials.
To alleviate this problem, we calculate the discrete value of $\textup{e}^{-\beta V^\text{ext} (\vec{r})}$ for a grid point (used in \eqabrev~\eqref{eq:sc2} below) as a spatial average of $\textup{e}^{-\beta V^\text{ext} (\vec{r}_i)}$ over a sub grid with 49 grid points per bin.

\subsection{Training with Local Windows vs.\ Training with whole Profiles}
The straightforward way to train the neural network is to directly pass the simulation density profiles to the neural network with parameters (weights) $\theta$. 
In the network, $c_1^\text{ML}(\theta)$ is calculated and 
compared to $c_1^\text{sim}$, the average loss per point $\mathcal{L}_2=(c_1^\text{ML} - c_1^\text{sim})^2$ is backpropagated and $\theta$ are updated. 
One batch of training data, which is processed between every adaption of the weights as a result of the loss, consists either of one density profile or a small set of 
density profiles. 
This is fully functional, but precision can be enhanced by using local density windows of minimal size $L_\text{min} \times L_\text{min}$ (dictated by the range of convolutions,
$L_\text{min}=2r/\Delta L'+1=81$) as proposed by Sammüller et al.\ \cite{sammuellerc1,sammuellerc1howto}.
These windows have to be chosen in an efficient way to optimize the batching. 
Let $\{\rho^{(1)},\dots,\rho^{(N)}\}$ be the set of density profiles (all from different simulations [superscript]) and 
$\{\rho^{(i)}_1,\dots,\rho^{(i)}_M\}$ the set of density windows [subscript] belonging to the $i$-th density profile.
The set of such density windows comprises all possible $L_\text{min} \times L_\text{min}$ squares from the simulation profile of size $L \times L$, where the corresponding $c_1$-value is finite.
For $L=200$, as in our case, this set comprises $M= 200\cdot200=40000$ windows (a few less if walls are present).

When creating a batch to forward to the neural network, random elements of the full list fo density windows
$\{\rho^{(1)}_1,\dots,\rho^{(1)}_M,\dots,\rho^{(N)}_1,\dots,\rho^{(N)}_M\}$ are chosen, e.g. $\{\rho^{(1)}_5,\rho^{(1)}_{23},\rho^{(3)}_7,\rho^{(4)}_4,\dots\}$, 
i.~e.\ a mixture of different windows from different density profiles. 
In contrast, if whole density profiles are used, we basically pass a batch of all density windows belonging to this density profile to the network 
(i.~e.\ $\{\rho^{(i)}_1,\dots,\rho^{(i)}_M\}$), and one cannot shuffle this for the next epoch. 
Moreover, batches get quite large, which is also not optimal \cite{batchsize1,batchsize2} and will, for a smoother discretization or the 3D case, become the main problem.

Density windows larger than $L_\text{min} \times L_\text{min}$
could be a good middle ground, however, minimal density windows containing only one relevant $c_1$-value already turned out to be sufficiently fast for training, and
therefore this option was not pursued further.
For our final training, we used a total of $N_\text{train}=53$ simulated profiles and a batch size of 96 minimal windows, which are randomly chosen from the full list of density windows.
Note that this batch size is much lower than the $M=40000$ minimal windows discussed before, which can be extracted from a single density profile.
The test set, generated similarly as the training set, contains $N_\text{test}= 24$ simulated profiles.

\section{Results}

\subsection{Self-consistent Density Profiles}
Training results in optimal weights $\theta^*$ and an associated ML map $c_1^*[\rho]=c_1^\text{ML}(\theta^*,[\rho])$. 
This can be used to generate self-consistent density profiles using \eqabrev~\eqref{eq:sc} in the form
\begin{equation}
	 \label{eq:sc2}
	 \rho_\text{ML} (\vec{r}) = \exp\left( c_1^* - \beta \, V^\text{ext} (\vec{r}) + \beta \, \mu \right)\;,
\end{equation}
\begin{figure}[h!]
    \centering
    \includegraphics[width=0.95\linewidth]{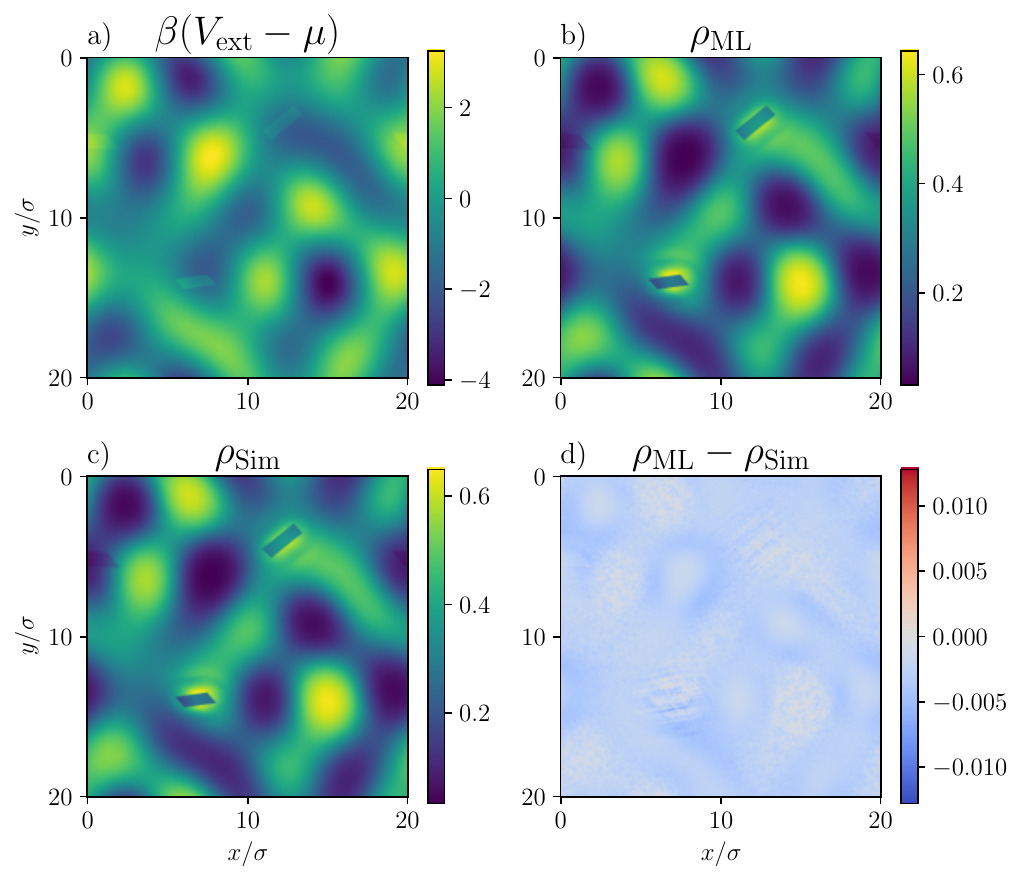}
	\caption{Typical example for a random potential (a), a self-consistent density profile obtained with the ML functional for this potential (b), the result of a GCMC simulation (c) and the difference between simulation and ML DFT result (d). Here, lateral system size is $2 \, L'$ (twice the training box size).}
    \label{fig:beyondthebox}
\end{figure}
which can be solved iteratively for external potentials within and outside the training set.
Here, the system size can be larger than the system size of training data. 
This is demonstrated in \figabrev~\ref{fig:beyondthebox} where we apply it to a simulation box of double the lateral size
and for a typical external potential of the form in \eqabrev~\eqref{eq:sine}, with additional plateaus. The average absolute error for the density is below 1\% of the average density
and the error itself is systematically negative which we interpret as the small finite-size error of our scheme (note that ML-DFT was trained on the smaller, $10\sigma\times 10\sigma$ system).

\begin{figure}[h!]
    \centering
    \includegraphics[width=0.95\linewidth]{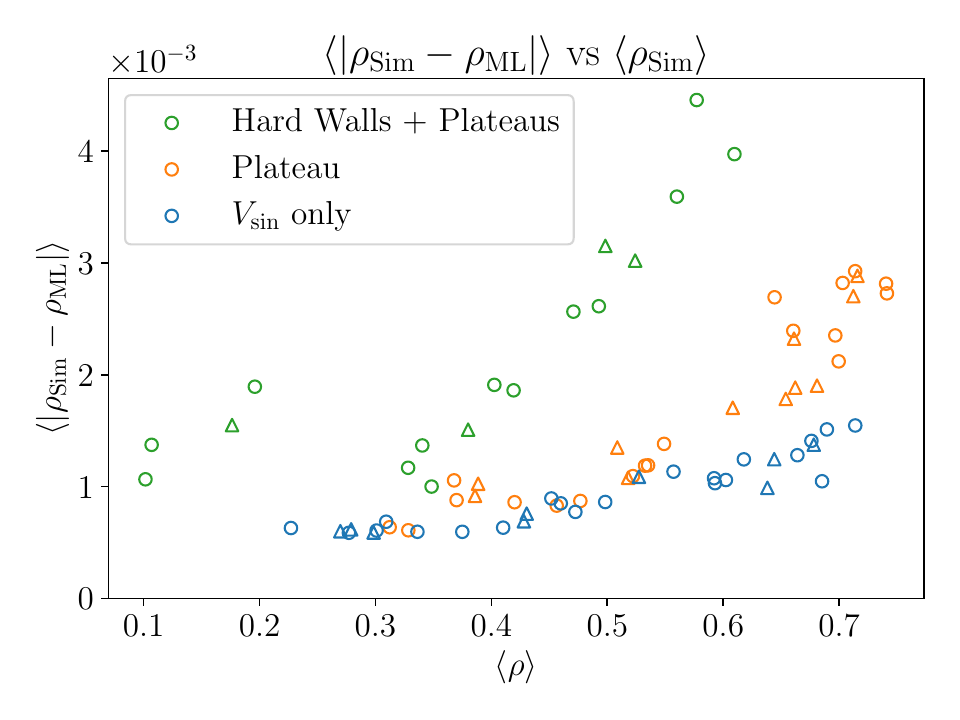}
    \caption{Performance overview of the neural network. The plot illustrates the mean absolute deviation between simulation and results obtained by 
	self-consistent iteration (\eqabrev~\eqref{eq:sc2}) as a function of the mean density. 
	The colors indicate the type of external potential (only sine functions in blue, a combination of sine functions and plateaus in orange and sine functions 
	with plateaus and hard walls in green). Circles are for systems of which some density windows have been used during training, triangles are for systems from the test set.
    }
    \label{fig:meanerror}
\end{figure}

In \figabrev~\ref{fig:meanerror}, the mean error per bin $\langle |\rho_\text{ML}-\rho_\text{sim}| \rangle $ is shown for some profiles from the training set 
(circles) and profiles from the test set (triangles) as a function of the mean density.  
There is no systematic difference visible between the two sets.
However,  it is visible that the network performs best for external potentials consisting only of smooth sine functions. 
For plateaus and hard walls, the network performs worse but still very well.
\begin{figure}[h!]
    \centering
    \includegraphics[width=0.95\linewidth]{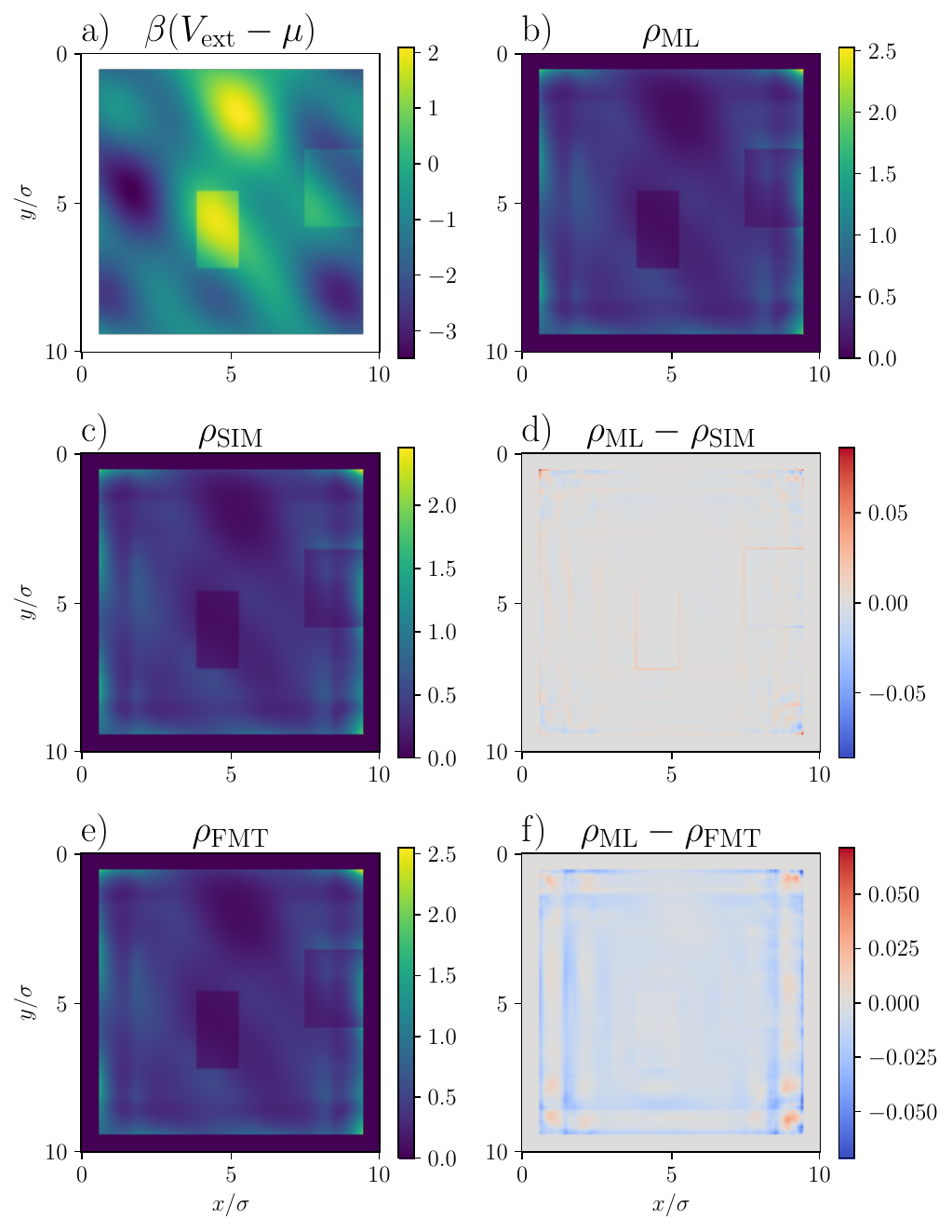}
    \caption{Typical example for an external potential with plateaus and hard walls (a) and the corresponding density profile obtained with the ML functional for this potential (b).
    This is compared to the result obtained by GCMC simulations (c,d) (where deviations at the edges of the plateaus and corners of the box are visible) and the FMT functional from \refabrev~\cite{martinsfunctional} with $a=11/4$, using the implementation of \refabrev~\cite{sam2018}) (e,f).}
    \label{fig:plateauproblems}
\end{figure}

The source of the increased error for plateau-type external potentials lies in comparably large density deviations on single bins at the edges of the plateaus due to a systematic discretization error, see \figabrev~\ref{fig:plateauproblems}. To obtain self-consistent profiles, $\beta V^\text{ext}(\vec{r})$ must be discretized in \eqabrev~\eqref{eq:sc2}, as already discussed in \secabrev~\ref{sec:discretization}. The method discussed there greatly reduces the error compared to simply choosing the potential value in the middle of the bin, but cannot fully compensate it, because the DFT code and the simulation code in the end use slightly different external potentials. The effect of this can be seen in \figabrev~\ref{fig:plateauproblems}, where the edges of the plateaus are still slightly visible when compared with simulation but not with FMT.
The error is further increased for density profiles which additionally feature hard walls at the box's edges. This can be attributed to very high densities near the corners of the box as seen in \figabrev~\ref{fig:plateauproblems}.
However, the agreement is still good. Deviations between ML and FMT are of similar magnitude as the deviations between ML and simulation.

\subsection{Pair Correlation Function}
The pair correlation function $g(r)$ can be obtained from the test particle route, i.e. solving \eqabrev~\eqref{eq:sc2} for an external potential corresponding to a fixed disk at the origin,
which gives $\rho(r) = \rho_0 \, g(r)$ where $\rho_0(\mu)$ is the bulk density for chemical potential $\mu$.
Here we set the external potential to infinity for all points further than half the particles' diameter away from the center of the simulation box, and zero otherwise. 
\begin{figure}[h!]
    \centering
    \includegraphics[width=0.95\linewidth]{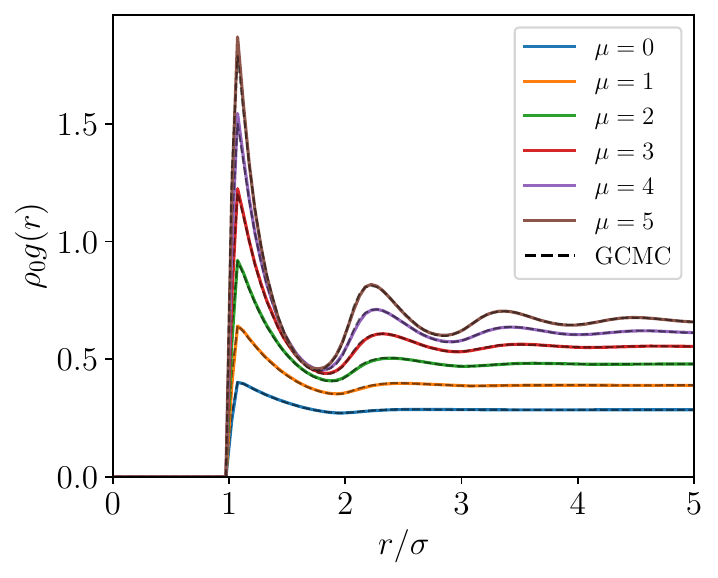}
    \caption{Comparison between $\rho_0 g(r)$ obtained by simulation (dashed) and by self-consistent Picard-Iteration using the neural network.
    Both profiles have been calculated by placing the disk potential in the middle of the numerical box, defining radial bins and averaging over those square density bins whose midpoints fall into a certain radial bin.}
    \label{fig:gr}
\end{figure}

\figabrev~\ref{fig:gr} compares the results for $\rho_0 g(r)$ from simulations and the ML functional. Good agreement is obtained, with some minor deviations in the first peak height which are most probably caused by the discretization problem discussed before. This is also a serious test for the extrapolation capabilities of the ML network, because the network was never trained with any radial potentials, furthermore, the only hard potentials encountered in the training were box walls in a rectilinear geometry. This appears to be promising for applications in arbitrary external potential geometries.

\subsection{Convolutional Kernels}
\label{sec:kernels}
\begin{figure}[h!]
    \centering
    \includegraphics[width=0.95\linewidth]{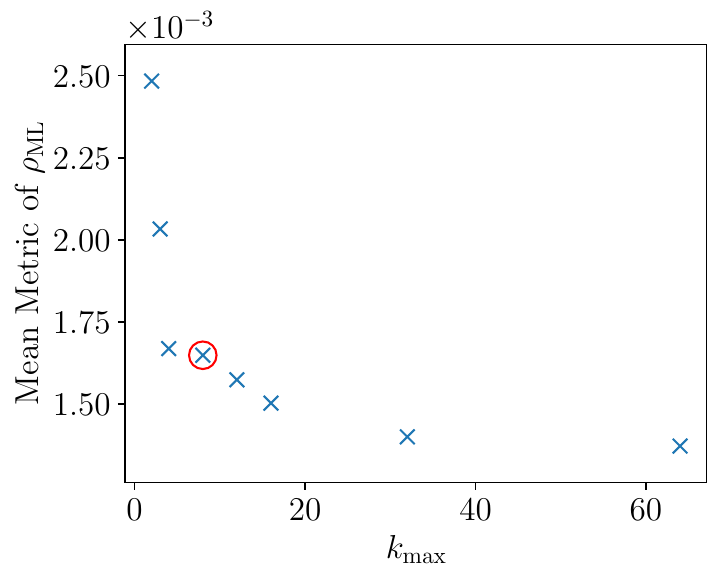}
	\caption{
    Mean absolute error $\langle | \rho_\text{ML}-\rho_\text{sim} | \rangle$ averaged over all density profiles as a function of the number of convolution kernels $k_\text{max}$. Already for very few kernels a good accuracy is reached. The highlighted value $k_\text{max} = 8$ is used throughout the paper.
	}
    \label{fig:lossvskernels}
\end{figure}

In \figabrev~\ref{fig:lossvskernels}, the mean metric of all profiles as a function of $\km$ (number of convolution kernels) is shown. 
Not unexpectedly, the model improves when adding more convolution kernels. As the metric is already very small for only a few kernels, the improvement is, however, also rather small. This justifies the value $\km = 8$ used throughout the paper which comes with a very low computational cost and is comparable to the number of FMT weights.

Since the ML network has been designed with a structure similar to FMT, it is interesting to study the optimized kernels $\omega_i(\vec{r})$ and $\tilde{\omega}_i(\vec{r})$.
In FMT (for one--component hard disks), these kernels (in the notation of \refabrev~\cite{martinsfunctional}) are given by 
\begin{eqnarray*}
	\omega^{(m)} = \underbrace{ \hat{\vec{r}} \, \otimes...\otimes \, \hat{\vec{r}} }_{m\;\text{times}} \, \delta(R-r)\;, \quad \omega_2=\theta(R-r)\;, 
\end{eqnarray*}	
where $R=\sigma/2$ and $\hat{\vec{r}}$ is a 2D unit vector. The kernels $\omega^{(m)}$ are generalized tensor weights ($m=0$ corresponds to a scalar kernel, 
$m=1$ to a vector kernel, $m=2$ to a second-rank tensor kernel). The functional derived in \refabrev~\cite{martinsfunctional} uses these tensor kernels up to $m=2$.  
The graphical representation of the scalar kernel $w_2$ is simply a homogeneously filled circle of radius $R$, whereas the tensor kernels are only non-zero on this circle and
exhibit a multipolar symmetry there: monopole for $m=0$, dipole for $m=1$ and quadrupole for $m=2$. As stated above, FMT imposes the restriction
$\tilde{\omega}_i(\vec{r})=\omega_i(\vec{-r})$.

\begin{figure}[h!]
    \centering
    \includegraphics[width=0.49\linewidth]{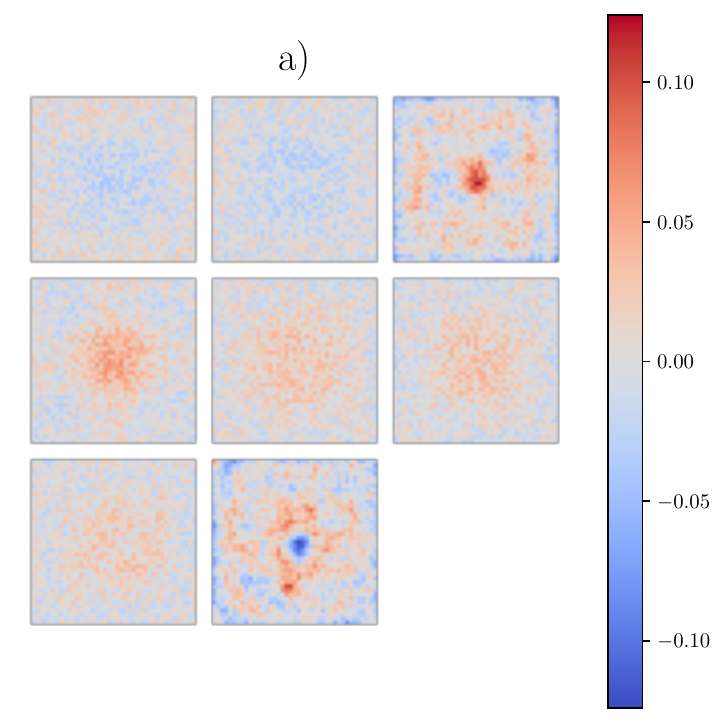}
    \includegraphics[width=0.49\linewidth]{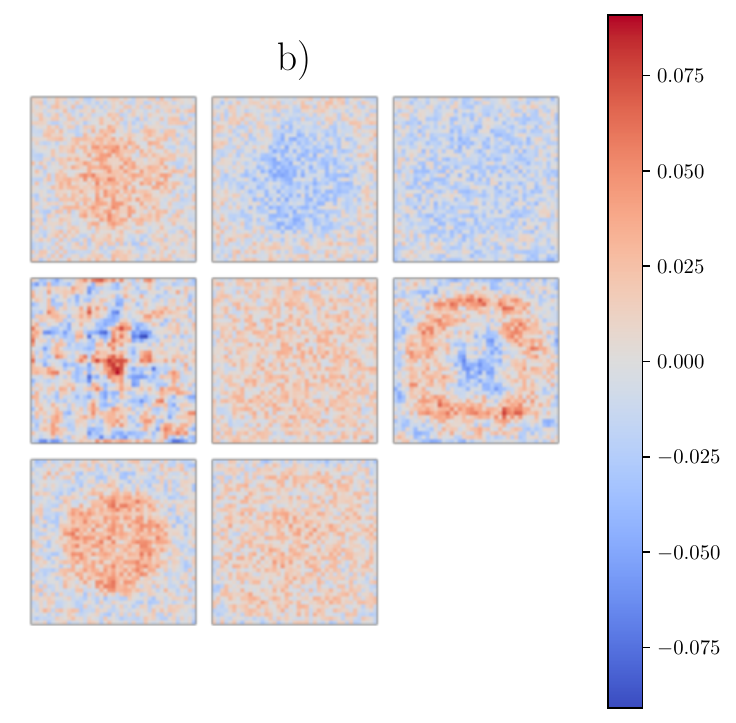}
    \caption{Convolution kernels $\omega_k$ (a) and $\tilde{\omega}_i$ (b) for the optimized ML functional with $\km=8$.
    }
    \label{fig:kernels}
\end{figure}

For the ML functional, the optimized kernels are shown in \figabrev~\ref{fig:kernels}.
Not unexpectedly, ML produces kernels that are in general non-zero over the whole kernel range which is a $r \times r$ square,
and there is no obvious link between $\tilde{\omega}_i$ and $\omega_i$. Quite a few kernels exhibit near-radial symmetry (are monopole-like), but there is no clear preponderance of dipole or quadrupole symmetry in the asymmetric kernels.
In fact, some kernels are very inhomgeneous with only random speckles.

\begin{figure}[h!]
    \centering
    \includegraphics[width=0.49\linewidth]{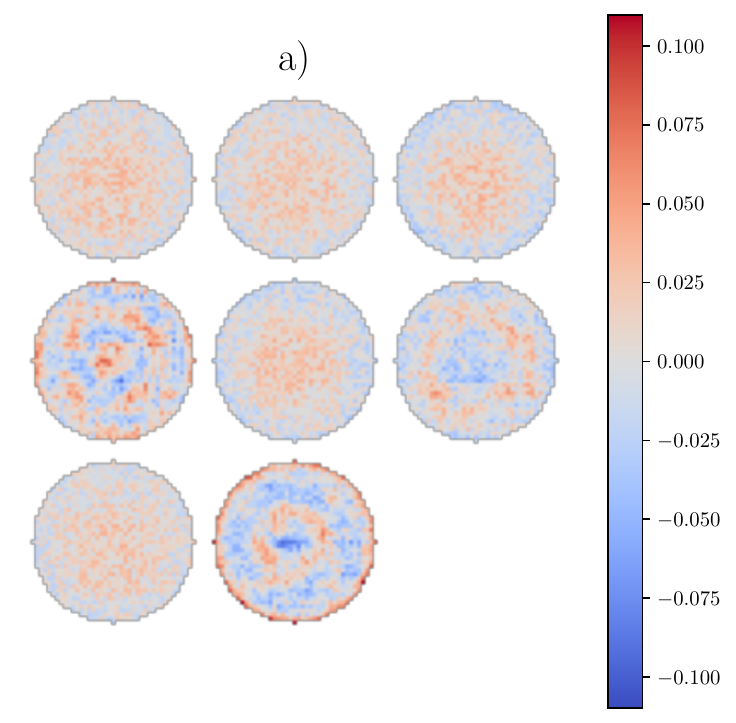}
    \includegraphics[width=0.49\linewidth]{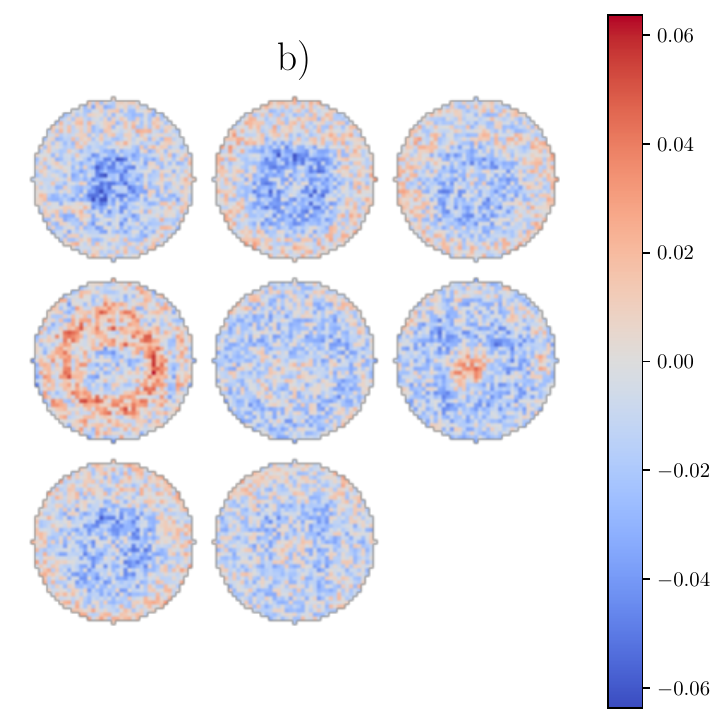}
    \caption{Convolution kernels $\omega_k$ (a) and $\tilde{\omega}_i$ (b) for an optimized ML functional with $\km=8$
    and an additional radial constraint on the kernels. }
    \label{fig:kernelsspherical}
\end{figure}

A radial confinement as suggested by FMT can also be enforced externally by setting all weights outside the disk to zero during training. The resulting functional shows a similar performance as before. \figabrev~\ref{fig:kernelsspherical} shows the results for the corresponding kernels. Interestingly, such confined kernels also exhibit a more radial structure in their interior, although this was not manually enforced.

\begin{figure}[h!]
    \centering
    \includegraphics[width=0.49\linewidth]{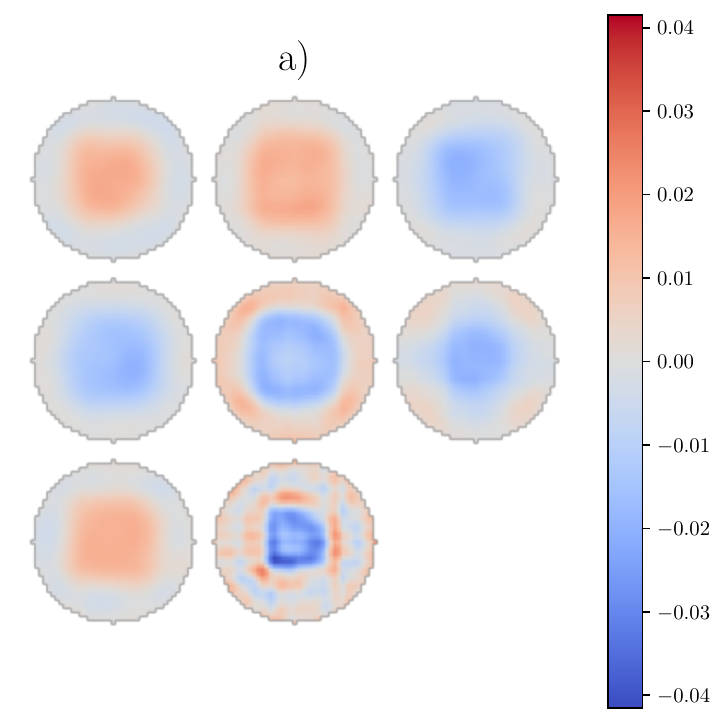}
    \includegraphics[width=0.49\linewidth]{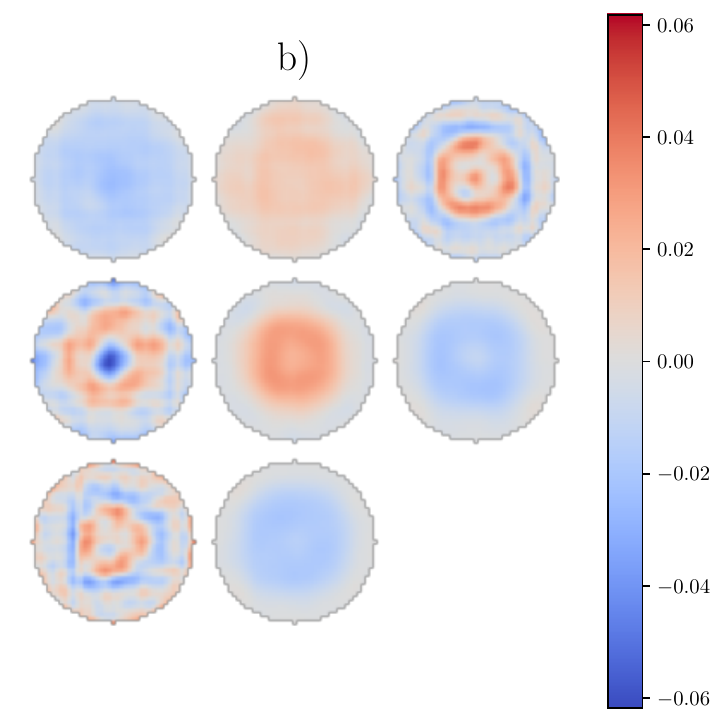}
    \caption{Convolution kernels $\omega_k$ (a) and $\tilde{\omega}_i$ (b) for an optimized ML functional with $\km=8$, an additional radial constraint on the kernels and slight smoothing during training.}
    \label{fig:kernelssphericalsmooth}
\end{figure}

The kernels resulting from the current training procedure look rather noisy. This noise is possibly not beneficial for the network accuracy outside of the training set and is probably a result of the limited quantity of training data, as well as the noise of the simulation.
To alleviate this problem, one can smooth the kernels during the training. Our method of choice to do so was applying a Gaussian blur with a standard deviation of $0.5$ to $1$ every 20 to 50 epochs.
Such a smoothing procedure does indeed improve the overall result slightly. However, the main advantage is that the reduction in the granularity of the kernels makes them more ``readable''. This is demonstrated in \figabrev~\ref{fig:kernelssphericalsmooth}.

\subsection{Comparison to the Model Designed by Sammüller et al.} \label{sec:sammuellernet}
\begin{figure}
    \centering
    \includegraphics{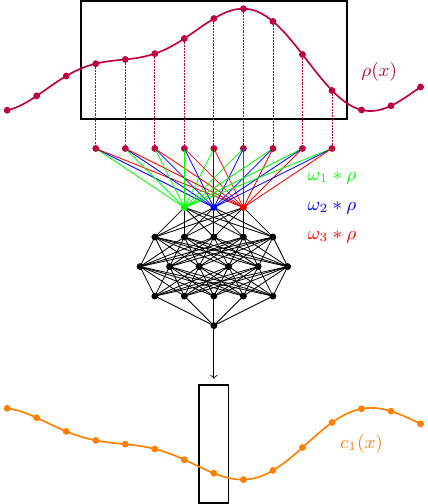}
    \caption{Visualization of the network introduced by Sammüller et al.\ in \cite{sammuellerc1,sammuellerc1howto} highlighting the ``manual convolution''. The number of layers and nodes per layer is adapted to match \figabrev~\ref{fig:convbayreuthnet}.}
    \label{fig:originalbayreuthnet}
\end{figure}
Local learning of $c_1$ was established by Sammüller et al. \cite{sammuellerc1,sammuellerc1howto} and our generalization to higher dimensions is heavily based on it. 
Here we briefly discuss the differences of our model to a straightforward 2D implementation of the Sammüller model.
In the Sammüller model, a fully connected network is fed with the density values of a local density window (minimal window) and produces a local output $c_1(\vec{r})$
at point $\vec{r}$ (see \figabrev~\ref{fig:originalbayreuthnet} for a schematics). To generate a full $c_1$-profile, the density profile is decomposed into local windows which are subsequently fed into the fully connected neural network to subsequently calculate $c_1$ on all points (see also the resources accompanying \refabrev~\cite{sammuellerc1howto}). This can be implemented in 2D straightforwardly and produces sensible results, but it is very slow when it comes to determining self-consistent density profiles using the iteration of \eqabrev~\eqref{eq:sc2}. 
This is due to the massive overhead of creating these local density windows and feeding them to the neural network individually. 
In 1D applications this is indeed not an issue for the computational time, but in 2D the time loss is significant.
We estimate that in 3D this straightforward method is not applicable anymore.

\begin{figure}[h!]
    \centering
    \includegraphics{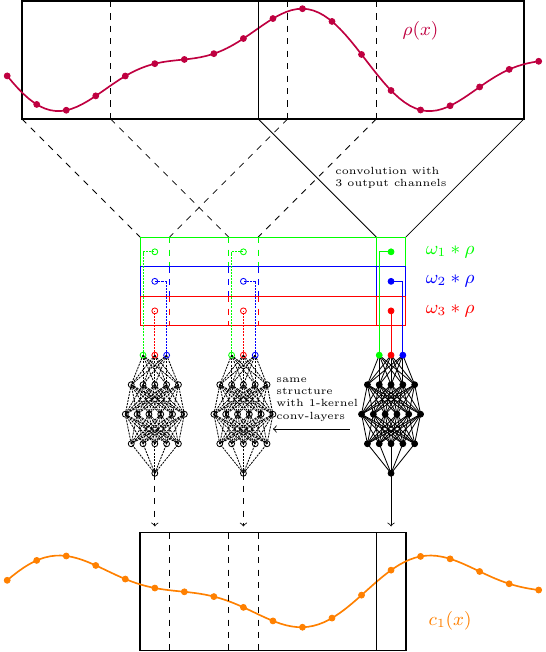}
    \caption{Schematic illustration of a convolutional implementation of the Sammüller network for a 1D geometry with $k_1=3$ convolution kernels. The figure illustrates the calculation of a profile of $c_1$ values. The parallel use of the same convolutional kernels is hinted with dashed lines. 
    When comparing this to \figabrev~\ref{fig:networkarchitecture}, note that the latter illustrates the generation of only a single $c_1$-value instead of a finite section of a $c_1$ profile here. 
    Furthermore the discretization is not comparable as range of the two convolutions $\omega_i$ and $\tilde{\omega}_i$ in \figabrev~\ref{fig:networkarchitecture} should be half of the convolution range here (to arrive at the same minimal window width).}
    \label{fig:convbayreuthnet}
\end{figure}

From \figabrev~\ref{fig:originalbayreuthnet} it is seen that the weights connecting the input layer and the first hidden layer basically implement a convolution kernel 
(more precisely, $k_1$ convolution kernels [channels] where $k_1$ is the number of nodes in the first hidden layer), producing $k_1$ weighted densities.
The following layer is only a function of the first hidden layer, i.~e.\ of the weighted densities, and therefore the whole network can, just as our network, 
be implemented by using convolutional layers only. 
The kernel size of the first convolution is given by the size of the input layer, while all other layers have kernels of size 1.
\figabrev~\ref{fig:convbayreuthnet} illustrates the convolutional implementation of the Sammüller network, it is fully equivalent in its action on an arbitrary 
density profile to the original network. In actual computations, the overhead mentioned above is eliminated, leading to a performance increase of at least a factor 20 (heavily dependent on the sophistication of the implementation).

\begin{figure}[t!]
    \centering
    \includegraphics[width=0.49\linewidth]{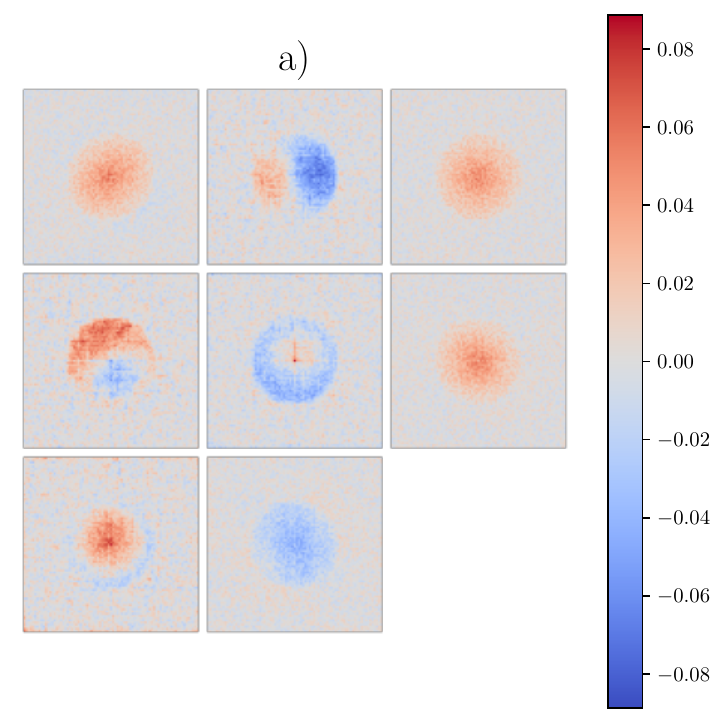}
    \includegraphics[width=0.49\linewidth]{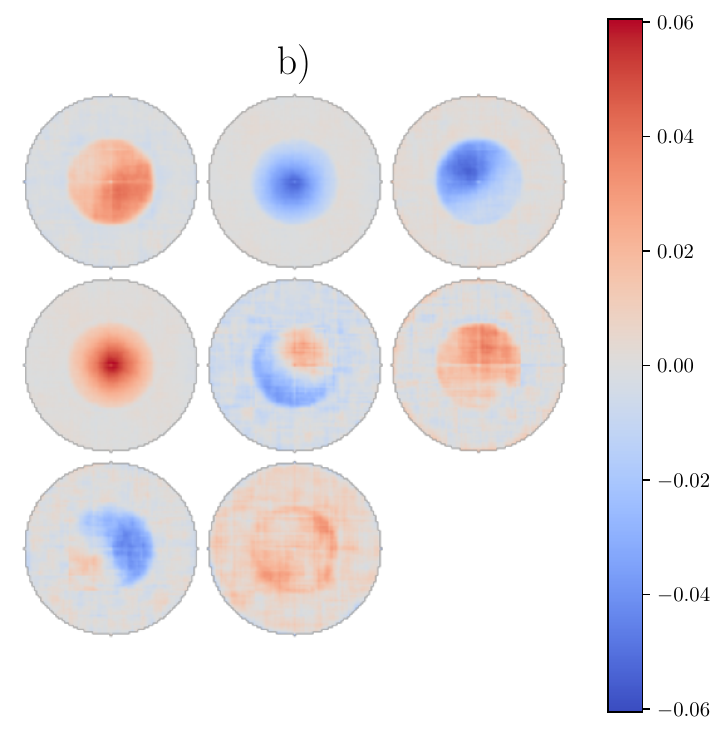}
    \caption{Convolution kernels $\omega_i$ for two 2D convolutional implementation of the Sammüller model. 
    a) kernels with square support. b) kernels with circular support and smoothing as in \figabrev~\ref{fig:kernelssphericalsmooth}. The network is required to have the same minimal window size as the implementation discussed before, resulting in doubling the width of the convolutional kernels.}
    \label{fig:kernelsbayreuth}
\end{figure}

We also benchmarked the loss for different values of $k_1$ for this model and although \refabrev~\cite{sammuellerc1howto} uses $k_1=128$ kernels, between 8 and 12 kernels are already sufficient to produce an accurate ML functional (but we also have more hidden layers).
Hence, with a convolutional implementation of the Sammüller network, one can also easily visualize its kernels, see \figabrev~\ref{fig:kernelsbayreuth}. The lateral extension of the kernels is taken to be 2$\sigma$ in order to have the same minimal window size as the FMT-like networks investigated before.
Surprisingly, the kernels have a very low absolute value outside the support of the FMT-like kernels.
As a distinct feature, some kernels show a marked dipolar symmetry which was not found in the FMT-like network.    
\FloatBarrier

\section{Conclusion and Outlook}
Based on the local learning method introduced by Sammüller et al.\ \cite{sammuellerc1,sammuellerc1howto}, we implemented a neural network for the hard disk system calculating $c_1 [\rho]$ with a structure inspired by FMT for a truly two-dimensional geometry. 
By using convolutional layers, we could demonstrate a drastically improved performance in 2D that should make the method fully applicable to 2D inhomogeneous systems with more general pair potentials beyond hard disks.
Also, we anticipate no conceptual or numerical bottlenecks in extending this method to fully inhomogeneous systems
in 3D besides the much increased, but doable additional effort in generating smooth density profiles.

\section*{Acknowledgments}
We thank Alessandro Simon for useful discussions. This work is supported by the Deutsche Forschungsgemeinschaft (DFG, German Research Foundation), project OE 285/9-1.

%\nocite{*}

\bibliography{lit}

\begin{thebibliography}{34}%
\makeatletter
\providecommand \@ifxundefined [1]{%
 \@ifx{#1\undefined}
}%
\providecommand \@ifnum [1]{%
 \ifnum #1\expandafter \@firstoftwo
 \else \expandafter \@secondoftwo
 \fi
}%
\providecommand \@ifx [1]{%
 \ifx #1\expandafter \@firstoftwo
 \else \expandafter \@secondoftwo
 \fi
}%
\providecommand \natexlab [1]{#1}%
\providecommand \enquote  [1]{``#1''}%
\providecommand \bibnamefont  [1]{#1}%
\providecommand \bibfnamefont [1]{#1}%
\providecommand \citenamefont [1]{#1}%
\providecommand \href@noop [0]{\@secondoftwo}%
\providecommand \href [0]{\begingroup \@sanitize@url \@href}%
\providecommand \@href[1]{\@@startlink{#1}\@@href}%
\providecommand \@@href[1]{\endgroup#1\@@endlink}%
\providecommand \@sanitize@url [0]{\catcode `\\12\catcode `\$12\catcode
  `\&12\catcode `\#12\catcode `\^12\catcode `\_12\catcode `\%12\relax}%
\providecommand \@@startlink[1]{}%
\providecommand \@@endlink[0]{}%
\providecommand \url  [0]{\begingroup\@sanitize@url \@url }%
\providecommand \@url [1]{\endgroup\@href {#1}{\urlprefix }}%
\providecommand \urlprefix  [0]{URL }%
\providecommand \Eprint [0]{\href }%
\providecommand \doibase [0]{https://doi.org/}%
\providecommand \selectlanguage [0]{\@gobble}%
\providecommand \bibinfo  [0]{\@secondoftwo}%
\providecommand \bibfield  [0]{\@secondoftwo}%
\providecommand \translation [1]{[#1]}%
\providecommand \BibitemOpen [0]{}%
\providecommand \bibitemStop [0]{}%
\providecommand \bibitemNoStop [0]{.\EOS\space}%
\providecommand \EOS [0]{\spacefactor3000\relax}%
\providecommand \BibitemShut  [1]{\csname bibitem#1\endcsname}%
\let\auto@bib@innerbib\@empty
%</preamble>
\bibitem [{\citenamefont {Evans}(1979)}]{bobstart}%
  \BibitemOpen
  \bibfield  {author} {\bibinfo {author} {\bibfnamefont {R.}~\bibnamefont
  {Evans}},\ }\bibfield  {title} {\bibinfo {title} {The nature of the
  liquid-vapour interface and other topics in the statistical mechanics of
  non-uniform, classical fluids},\ }\href
  {https://doi.org/10.1080/00018737900101365} {\bibfield  {journal} {\bibinfo
  {journal} {Adv. Phys.}\ }\textbf {\bibinfo {volume} {28}},\ \bibinfo {pages}
  {143} (\bibinfo {year} {1979})}\BibitemShut {NoStop}%
\bibitem [{\citenamefont {Roth}(2010)}]{rothsreview}%
  \BibitemOpen
  \bibfield  {author} {\bibinfo {author} {\bibfnamefont {R.}~\bibnamefont
  {Roth}},\ }\bibfield  {title} {\bibinfo {title} {Fundamental measure theory
  for hard-sphere mixtures: a review},\ }\href
  {https://doi.org/10.1088/0953-8984/22/6/063102} {\bibfield  {journal}
  {\bibinfo  {journal} {J. Phys.: Condens. Matter}\ }\textbf {\bibinfo {volume}
  {22}},\ \bibinfo {pages} {063102} (\bibinfo {year} {2010})}\BibitemShut
  {NoStop}%
\bibitem [{\citenamefont {Hansen-Goos}\ and\ \citenamefont
  {Mecke}(2009)}]{mecke2009}%
  \BibitemOpen
  \bibfield  {author} {\bibinfo {author} {\bibfnamefont {H.}~\bibnamefont
  {Hansen-Goos}}\ and\ \bibinfo {author} {\bibfnamefont {K.}~\bibnamefont
  {Mecke}},\ }\bibfield  {title} {\bibinfo {title} {Fundamental measure theory
  for inhomogeneous fluids of nonspherical hard particles},\ }\href
  {https://doi.org/10.1103/PhysRevLett.102.018302} {\bibfield  {journal}
  {\bibinfo  {journal} {Phys. Rev. Lett.}\ }\textbf {\bibinfo {volume} {102}},\
  \bibinfo {pages} {018302} (\bibinfo {year} {2009})}\BibitemShut {NoStop}%
\bibitem [{\citenamefont {Wittmann}\ \emph {et~al.}(2017)\citenamefont
  {Wittmann}, \citenamefont {Sitta}, \citenamefont {Smallenburg},\ and\
  \citenamefont {L\"owen}}]{wittmann2017}%
  \BibitemOpen
  \bibfield  {author} {\bibinfo {author} {\bibfnamefont {R.}~\bibnamefont
  {Wittmann}}, \bibinfo {author} {\bibfnamefont {C.~E.}\ \bibnamefont {Sitta}},
  \bibinfo {author} {\bibfnamefont {F.}~\bibnamefont {Smallenburg}},\ and\
  \bibinfo {author} {\bibfnamefont {H.}~\bibnamefont {L\"owen}},\ }\bibfield
  {title} {\bibinfo {title} {Phase diagram of two-dimensional hard rods from
  fundamental mixed measure density functional theory},\ }\href
  {https://doi.org/10.1063/1.4996131} {\bibfield  {journal} {\bibinfo
  {journal} {J. Chem. Phys.}\ }\textbf {\bibinfo {volume} {147}},\ \bibinfo
  {pages} {134908} (\bibinfo {year} {2017})}\BibitemShut {NoStop}%
\bibitem [{\citenamefont {Lutsko}(2020)}]{lutsko2020}%
  \BibitemOpen
  \bibfield  {author} {\bibinfo {author} {\bibfnamefont {J.~F.}\ \bibnamefont
  {Lutsko}},\ }\bibfield  {title} {\bibinfo {title} {Explicitly stable
  fundamental-measure-theory models for classical density functional theory},\
  }\href {https://doi.org/10.1103/PhysRevE.102.062137} {\bibfield  {journal}
  {\bibinfo  {journal} {Phys. Rev. E}\ }\textbf {\bibinfo {volume} {102}},\
  \bibinfo {pages} {062137} (\bibinfo {year} {2020})}\BibitemShut {NoStop}%
\bibitem [{\citenamefont {Stopper}\ \emph {et~al.}(2018)\citenamefont
  {Stopper}, \citenamefont {Hirschmann}, \citenamefont {Oettel},\ and\
  \citenamefont {Roth}}]{stopper2018}%
  \BibitemOpen
  \bibfield  {author} {\bibinfo {author} {\bibfnamefont {D.}~\bibnamefont
  {Stopper}}, \bibinfo {author} {\bibfnamefont {F.}~\bibnamefont {Hirschmann}},
  \bibinfo {author} {\bibfnamefont {M.}~\bibnamefont {Oettel}},\ and\ \bibinfo
  {author} {\bibfnamefont {R.}~\bibnamefont {Roth}},\ }\bibfield  {title}
  {\bibinfo {title} {Bulk structural information from density functionals for
  patchy particles},\ }\href {https://doi.org/10.1063/1.5064780} {\bibfield
  {journal} {\bibinfo  {journal} {J. Chem. Phys.}\ }\textbf {\bibinfo {volume}
  {149}},\ \bibinfo {pages} {224503} (\bibinfo {year} {2018})}\BibitemShut
  {NoStop}%
\bibitem [{\citenamefont {Barthes}\ \emph {et~al.}(2024)\citenamefont
  {Barthes}, \citenamefont {Bernet}, \citenamefont {Grégoire},\ and\
  \citenamefont {Miqueu}}]{barthes2024}%
  \BibitemOpen
  \bibfield  {author} {\bibinfo {author} {\bibfnamefont {A.}~\bibnamefont
  {Barthes}}, \bibinfo {author} {\bibfnamefont {T.}~\bibnamefont {Bernet}},
  \bibinfo {author} {\bibfnamefont {D.}~\bibnamefont {Grégoire}},\ and\
  \bibinfo {author} {\bibfnamefont {C.}~\bibnamefont {Miqueu}},\ }\bibfield
  {title} {\bibinfo {title} {A molecular density functional theory for
  associating fluids in 3d geometries},\ }\href
  {https://doi.org/10.1063/5.0180795} {\bibfield  {journal} {\bibinfo
  {journal} {J. Chem. Phys.}\ }\textbf {\bibinfo {volume} {160}},\ \bibinfo
  {pages} {054704} (\bibinfo {year} {2024})}\BibitemShut {NoStop}%
\bibitem [{\citenamefont {Zimmermann}\ and\ \citenamefont
  {Oettel}(2024)}]{michaelspaper}%
  \BibitemOpen
  \bibfield  {author} {\bibinfo {author} {\bibfnamefont {M.}~\bibnamefont
  {Zimmermann}}\ and\ \bibinfo {author} {\bibfnamefont {M.}~\bibnamefont
  {Oettel}},\ }\bibfield  {title} {\bibinfo {title} {Lattice fundamental
  measure theory beyond 0d cavities: Dimers on square lattices},\ }\bibfield
  {journal} {\bibinfo  {journal} {J. Stat. Phys.}\ }\textbf {\bibinfo {volume}
  {191}},\ \href {https://doi.org/10.1007/s10955-024-03350-4}
  {10.1007/s10955-024-03350-4} (\bibinfo {year} {2024})\BibitemShut {NoStop}%
\bibitem [{\citenamefont {G\"ul}\ \emph {et~al.}(2024)\citenamefont {G\"ul},
  \citenamefont {Roth},\ and\ \citenamefont {Evans}}]{melihspaper}%
  \BibitemOpen
  \bibfield  {author} {\bibinfo {author} {\bibfnamefont {M.}~\bibnamefont
  {G\"ul}}, \bibinfo {author} {\bibfnamefont {R.}~\bibnamefont {Roth}},\ and\
  \bibinfo {author} {\bibfnamefont {R.}~\bibnamefont {Evans}},\ }\bibfield
  {title} {\bibinfo {title} {Using test particle sum rules to construct
  accurate functionals in classical density functional theory},\ }\href
  {https://doi.org/10.1103/PhysRevE.110.064115} {\bibfield  {journal} {\bibinfo
   {journal} {Phys. Rev. E}\ }\textbf {\bibinfo {volume} {110}},\ \bibinfo
  {pages} {064115} (\bibinfo {year} {2024})}\BibitemShut {NoStop}%
\bibitem [{\citenamefont {Lin}\ and\ \citenamefont {Oettel}(2019)}]{sammldft}%
  \BibitemOpen
  \bibfield  {author} {\bibinfo {author} {\bibfnamefont {S.-C.}\ \bibnamefont
  {Lin}}\ and\ \bibinfo {author} {\bibfnamefont {M.}~\bibnamefont {Oettel}},\
  }\bibfield  {title} {\bibinfo {title} {{A classical density functional from
  machine learning and a convolutional neural network}},\ }\href
  {https://doi.org/10.21468/SciPostPhys.6.2.025} {\bibfield  {journal}
  {\bibinfo  {journal} {SciPost Phys.}\ }\textbf {\bibinfo {volume} {6}},\
  \bibinfo {pages} {025} (\bibinfo {year} {2019})}\BibitemShut {NoStop}%
\bibitem [{\citenamefont {Lin}\ \emph {et~al.}(2020)\citenamefont {Lin},
  \citenamefont {Martius},\ and\ \citenamefont {Oettel}}]{sammldft2}%
  \BibitemOpen
  \bibfield  {author} {\bibinfo {author} {\bibfnamefont {S.-C.}\ \bibnamefont
  {Lin}}, \bibinfo {author} {\bibfnamefont {G.}~\bibnamefont {Martius}},\ and\
  \bibinfo {author} {\bibfnamefont {M.}~\bibnamefont {Oettel}},\ }\bibfield
  {title} {\bibinfo {title} {{Analytical classical density functionals from an
  equation learning network}},\ }\href {https://doi.org/10.1063/1.5135919}
  {\bibfield  {journal} {\bibinfo  {journal} {J. Chem. Phys.}\ }\textbf
  {\bibinfo {volume} {152}},\ \bibinfo {pages} {021102} (\bibinfo {year}
  {2020})}\BibitemShut {NoStop}%
\bibitem [{\citenamefont {Cats}\ \emph {et~al.}(2021)\citenamefont {Cats},
  \citenamefont {Kuipers}, \citenamefont {de~Wind}, \citenamefont {van Damme},
  \citenamefont {Coli}, \citenamefont {Dijkstra},\ and\ \citenamefont {van
  Roij}}]{catsroij}%
  \BibitemOpen
  \bibfield  {author} {\bibinfo {author} {\bibfnamefont {P.}~\bibnamefont
  {Cats}}, \bibinfo {author} {\bibfnamefont {S.}~\bibnamefont {Kuipers}},
  \bibinfo {author} {\bibfnamefont {S.}~\bibnamefont {de~Wind}}, \bibinfo
  {author} {\bibfnamefont {R.}~\bibnamefont {van Damme}}, \bibinfo {author}
  {\bibfnamefont {G.~M.}\ \bibnamefont {Coli}}, \bibinfo {author}
  {\bibfnamefont {M.}~\bibnamefont {Dijkstra}},\ and\ \bibinfo {author}
  {\bibfnamefont {R.}~\bibnamefont {van Roij}},\ }\bibfield  {title} {\bibinfo
  {title} {Machine-learning free-energy functionals using density profiles from
  simulations},\ }\href {https://doi.org/10.1063/5.0042558} {\bibfield
  {journal} {\bibinfo  {journal} {APL Mater.}\ }\textbf {\bibinfo {volume}
  {9}},\ \bibinfo {pages} {031109} (\bibinfo {year} {2021})}\BibitemShut
  {NoStop}%
\bibitem [{\citenamefont {Simon}\ \emph {et~al.}(2024)\citenamefont {Simon},
  \citenamefont {Weimar}, \citenamefont {Martius},\ and\ \citenamefont
  {Oettel}}]{alessandropatchyparticles}%
  \BibitemOpen
  \bibfield  {author} {\bibinfo {author} {\bibfnamefont {A.}~\bibnamefont
  {Simon}}, \bibinfo {author} {\bibfnamefont {J.}~\bibnamefont {Weimar}},
  \bibinfo {author} {\bibfnamefont {G.}~\bibnamefont {Martius}},\ and\ \bibinfo
  {author} {\bibfnamefont {M.}~\bibnamefont {Oettel}},\ }\bibfield  {title}
  {\bibinfo {title} {Machine learning of a density functional for anisotropic
  patchy particles},\ }\href {https://doi.org/10.1021/acs.jctc.3c01238}
  {\bibfield  {journal} {\bibinfo  {journal} {J. Chem. Theory Comput.}\
  }\textbf {\bibinfo {volume} {20}},\ \bibinfo {pages} {1062} (\bibinfo {year}
  {2024})}\BibitemShut {NoStop}%
\bibitem [{\citenamefont {Simon}\ \emph {et~al.}(2025)\citenamefont {Simon},
  \citenamefont {Belloni}, \citenamefont {Borgis},\ and\ \citenamefont
  {Oettel}}]{alessandropatchyparticles2}%
  \BibitemOpen
  \bibfield  {author} {\bibinfo {author} {\bibfnamefont {A.}~\bibnamefont
  {Simon}}, \bibinfo {author} {\bibfnamefont {L.}~\bibnamefont {Belloni}},
  \bibinfo {author} {\bibfnamefont {D.}~\bibnamefont {Borgis}},\ and\ \bibinfo
  {author} {\bibfnamefont {M.}~\bibnamefont {Oettel}},\ }\bibfield  {title}
  {\bibinfo {title} {The orientational structure of a model patchy particle
  fluid: Simulations, integral equations, density functional theory, and
  machine learning},\ }\href {https://doi.org/10.1063/5.0248694} {\bibfield
  {journal} {\bibinfo  {journal} {J. Chem. Phys.}\ }\textbf {\bibinfo {volume}
  {162}},\ \bibinfo {pages} {034503} (\bibinfo {year} {2025})}\BibitemShut
  {NoStop}%
\bibitem [{\citenamefont {Dijkman}\ \emph {et~al.}(2025)\citenamefont
  {Dijkman}, \citenamefont {Dijkstra}, \citenamefont {van Roij}, \citenamefont
  {Welling}, \citenamefont {van~de Meent},\ and\ \citenamefont
  {Ensing}}]{dftmlpaircorrmatching}%
  \BibitemOpen
  \bibfield  {author} {\bibinfo {author} {\bibfnamefont {J.}~\bibnamefont
  {Dijkman}}, \bibinfo {author} {\bibfnamefont {M.}~\bibnamefont {Dijkstra}},
  \bibinfo {author} {\bibfnamefont {R.}~\bibnamefont {van Roij}}, \bibinfo
  {author} {\bibfnamefont {M.}~\bibnamefont {Welling}}, \bibinfo {author}
  {\bibfnamefont {J.-W.}\ \bibnamefont {van~de Meent}},\ and\ \bibinfo {author}
  {\bibfnamefont {B.}~\bibnamefont {Ensing}},\ }\bibfield  {title} {\bibinfo
  {title} {Learning neural free-energy functionals with pair-correlation
  matching},\ }\href {https://doi.org/10.1103/PhysRevLett.134.056103}
  {\bibfield  {journal} {\bibinfo  {journal} {Phys. Rev. Lett.}\ }\textbf
  {\bibinfo {volume} {134}},\ \bibinfo {pages} {056103} (\bibinfo {year}
  {2025})}\BibitemShut {NoStop}%
\bibitem [{\citenamefont {Kelley}\ \emph {et~al.}(2024)\citenamefont {Kelley},
  \citenamefont {Quinton}, \citenamefont {Fazel}, \citenamefont {Karimitari},
  \citenamefont {Sutton},\ and\ \citenamefont {Sundararaman}}]{kelley2024}%
  \BibitemOpen
  \bibfield  {author} {\bibinfo {author} {\bibfnamefont {M.~M.}\ \bibnamefont
  {Kelley}}, \bibinfo {author} {\bibfnamefont {J.}~\bibnamefont {Quinton}},
  \bibinfo {author} {\bibfnamefont {K.}~\bibnamefont {Fazel}}, \bibinfo
  {author} {\bibfnamefont {N.}~\bibnamefont {Karimitari}}, \bibinfo {author}
  {\bibfnamefont {C.}~\bibnamefont {Sutton}},\ and\ \bibinfo {author}
  {\bibfnamefont {R.}~\bibnamefont {Sundararaman}},\ }\bibfield  {title}
  {\bibinfo {title} {Bridging electronic and classical density-functional
  theory using universal machine-learned functional approximations},\ }\href
  {https://doi.org/10.1063/5.0223792} {\bibfield  {journal} {\bibinfo
  {journal} {The Journal of Chemical Physics}\ }\textbf {\bibinfo {volume}
  {161}},\ \bibinfo {pages} {144101} (\bibinfo {year} {2024})}\BibitemShut
  {NoStop}%
\bibitem [{\citenamefont {Samm\"uller}\ and\ \citenamefont
  {Schmidt}(2024)}]{sammuellerpaircorrresponse}%
  \BibitemOpen
  \bibfield  {author} {\bibinfo {author} {\bibfnamefont {F.}~\bibnamefont
  {Samm\"uller}}\ and\ \bibinfo {author} {\bibfnamefont {M.}~\bibnamefont
  {Schmidt}},\ }\bibfield  {title} {\bibinfo {title} {Neural density
  functionals: Local learning and pair-correlation matching},\ }\href
  {https://doi.org/10.1103/PhysRevE.110.L032601} {\bibfield  {journal}
  {\bibinfo  {journal} {Phys. Rev. E}\ }\textbf {\bibinfo {volume} {110}},\
  \bibinfo {pages} {L032601} (\bibinfo {year} {2024})}\BibitemShut {NoStop}%
\bibitem [{\citenamefont {Sammüller}\ \emph {et~al.}(2023)\citenamefont
  {Sammüller}, \citenamefont {Hermann}, \citenamefont {de~las Heras},\ and\
  \citenamefont {Schmidt}}]{sammuellerc1}%
  \BibitemOpen
  \bibfield  {author} {\bibinfo {author} {\bibfnamefont {F.}~\bibnamefont
  {Sammüller}}, \bibinfo {author} {\bibfnamefont {S.}~\bibnamefont {Hermann}},
  \bibinfo {author} {\bibfnamefont {D.}~\bibnamefont {de~las Heras}},\ and\
  \bibinfo {author} {\bibfnamefont {M.}~\bibnamefont {Schmidt}},\ }\bibfield
  {title} {\bibinfo {title} {Neural functional theory for inhomogeneous fluids:
  Fundamentals and applications},\ }\href
  {https://doi.org/10.1073/pnas.2312484120} {\bibfield  {journal} {\bibinfo
  {journal} {Proc. Natl. Acad. Sci.}\ }\textbf {\bibinfo {volume} {120}},\
  \bibinfo {pages} {e2312484120} (\bibinfo {year} {2023})}\BibitemShut
  {NoStop}%
\bibitem [{\citenamefont {Sammüller}\ \emph {et~al.}(2024)\citenamefont
  {Sammüller}, \citenamefont {Hermann},\ and\ \citenamefont
  {Schmidt}}]{sammuellerc1howto}%
  \BibitemOpen
  \bibfield  {author} {\bibinfo {author} {\bibfnamefont {F.}~\bibnamefont
  {Sammüller}}, \bibinfo {author} {\bibfnamefont {S.}~\bibnamefont
  {Hermann}},\ and\ \bibinfo {author} {\bibfnamefont {M.}~\bibnamefont
  {Schmidt}},\ }\bibfield  {title} {\bibinfo {title} {Why neural functionals
  suit statistical mechanics},\ }\href
  {https://doi.org/10.1088/1361-648X/ad326f} {\bibfield  {journal} {\bibinfo
  {journal} {J. Phys.: Condens. Matter}\ }\textbf {\bibinfo {volume} {36}},\
  \bibinfo {pages} {243002} (\bibinfo {year} {2024})}\BibitemShut {NoStop}%
\bibitem [{\citenamefont {Samm\"uller}\ \emph {et~al.}(2025)\citenamefont
  {Samm\"uller}, \citenamefont {Schmidt},\ and\ \citenamefont
  {Evans}}]{sammuellerc1temperature}%
  \BibitemOpen
  \bibfield  {author} {\bibinfo {author} {\bibfnamefont {F.}~\bibnamefont
  {Samm\"uller}}, \bibinfo {author} {\bibfnamefont {M.}~\bibnamefont
  {Schmidt}},\ and\ \bibinfo {author} {\bibfnamefont {R.}~\bibnamefont
  {Evans}},\ }\bibfield  {title} {\bibinfo {title} {Neural density functional
  theory of liquid-gas phase coexistence},\ }\href
  {https://doi.org/10.1103/PhysRevX.15.011013} {\bibfield  {journal} {\bibinfo
  {journal} {Phys. Rev. X}\ }\textbf {\bibinfo {volume} {15}},\ \bibinfo
  {pages} {011013} (\bibinfo {year} {2025})}\BibitemShut {NoStop}%
\bibitem [{\citenamefont {Kampa}\ \emph {et~al.}(2025)\citenamefont {Kampa},
  \citenamefont {Samm\"uller}, \citenamefont {Schmidt},\ and\ \citenamefont
  {Evans}}]{kampapairpotentials}%
  \BibitemOpen
  \bibfield  {author} {\bibinfo {author} {\bibfnamefont {S.~M.}\ \bibnamefont
  {Kampa}}, \bibinfo {author} {\bibfnamefont {F.}~\bibnamefont {Samm\"uller}},
  \bibinfo {author} {\bibfnamefont {M.}~\bibnamefont {Schmidt}},\ and\ \bibinfo
  {author} {\bibfnamefont {R.}~\bibnamefont {Evans}},\ }\bibfield  {title}
  {\bibinfo {title} {Metadensity functional theory for classical fluids:
  Extracting the pair potential},\ }\href
  {https://doi.org/10.1103/PhysRevLett.134.107301} {\bibfield  {journal}
  {\bibinfo  {journal} {Phys. Rev. Lett.}\ }\textbf {\bibinfo {volume} {134}},\
  \bibinfo {pages} {107301} (\bibinfo {year} {2025})}\BibitemShut {NoStop}%
\bibitem [{\citenamefont {Samm\"uller}\ \emph {et~al.}(2024)\citenamefont
  {Samm\"uller}, \citenamefont {Robitschko}, \citenamefont {Hermann},\ and\
  \citenamefont {Schmidt}}]{sammuellerhyperdensity}%
  \BibitemOpen
  \bibfield  {author} {\bibinfo {author} {\bibfnamefont {F.}~\bibnamefont
  {Samm\"uller}}, \bibinfo {author} {\bibfnamefont {S.}~\bibnamefont
  {Robitschko}}, \bibinfo {author} {\bibfnamefont {S.}~\bibnamefont
  {Hermann}},\ and\ \bibinfo {author} {\bibfnamefont {M.}~\bibnamefont
  {Schmidt}},\ }\bibfield  {title} {\bibinfo {title} {Hyperdensity functional
  theory of soft matter},\ }\href
  {https://doi.org/10.1103/PhysRevLett.133.098201} {\bibfield  {journal}
  {\bibinfo  {journal} {Phys. Rev. Lett.}\ }\textbf {\bibinfo {volume} {133}},\
  \bibinfo {pages} {098201} (\bibinfo {year} {2024})}\BibitemShut {NoStop}%
\bibitem [{\citenamefont {Sammüller}\ and\ \citenamefont
  {Schmidt}(2024)}]{sammuellerhyperdensityhowto}%
  \BibitemOpen
  \bibfield  {author} {\bibinfo {author} {\bibfnamefont {F.}~\bibnamefont
  {Sammüller}}\ and\ \bibinfo {author} {\bibfnamefont {M.}~\bibnamefont
  {Schmidt}},\ }\bibfield  {title} {\bibinfo {title} {Why hyperdensity
  functionals describe any equilibrium observable},\ }\href
  {https://doi.org/10.1088/1361-648X/ad98da} {\bibfield  {journal} {\bibinfo
  {journal} {J. Phys.: Condens. Matter}\ }\textbf {\bibinfo {volume} {37}},\
  \bibinfo {pages} {083001} (\bibinfo {year} {2024})}\BibitemShut {NoStop}%
\bibitem [{\citenamefont {Bui}\ and\ \citenamefont
  {Cox}(2024)}]{buiionicfluidsc1}%
  \BibitemOpen
  \bibfield  {author} {\bibinfo {author} {\bibfnamefont {A.~T.}\ \bibnamefont
  {Bui}}\ and\ \bibinfo {author} {\bibfnamefont {S.~J.}\ \bibnamefont {Cox}},\
  }\href {https://arxiv.org/abs/2410.02556} {\bibinfo {title} {Learning
  classical density functionals for ionic fluids}} (\bibinfo {year} {2024}),\
  \Eprint {https://arxiv.org/abs/2410.02556} {arXiv:2410.02556
  [cond-mat.stat-mech]} \BibitemShut {NoStop}%
\bibitem [{\citenamefont {Yang}\ \emph {et~al.}(2024)\citenamefont {Yang},
  \citenamefont {Pan}, \citenamefont {Sun},\ and\ \citenamefont
  {Wu}}]{wuanisoc1}%
  \BibitemOpen
  \bibfield  {author} {\bibinfo {author} {\bibfnamefont {J.}~\bibnamefont
  {Yang}}, \bibinfo {author} {\bibfnamefont {R.}~\bibnamefont {Pan}}, \bibinfo
  {author} {\bibfnamefont {J.}~\bibnamefont {Sun}},\ and\ \bibinfo {author}
  {\bibfnamefont {J.}~\bibnamefont {Wu}},\ }\href
  {https://arxiv.org/abs/2411.03698} {\bibinfo {title} {High-dimensional
  operator learning for molecular density functional theory}} (\bibinfo {year}
  {2024}),\ \Eprint {https://arxiv.org/abs/2411.03698} {arXiv:2411.03698
  [physics.chem-ph]} \BibitemShut {NoStop}%
\bibitem [{\citenamefont {Simon}\ and\ \citenamefont
  {Oettel}(2024)}]{reviewalessandromartin}%
  \BibitemOpen
  \bibfield  {author} {\bibinfo {author} {\bibfnamefont {A.}~\bibnamefont
  {Simon}}\ and\ \bibinfo {author} {\bibfnamefont {M.}~\bibnamefont {Oettel}},\
  }\href {https://arxiv.org/abs/2406.07345} {\bibinfo {title} {Machine learning
  approaches to classical density functional theory}} (\bibinfo {year}
  {2024}),\ \Eprint {https://arxiv.org/abs/2406.07345} {arXiv:2406.07345
  [cond-mat.stat-mech]} \BibitemShut {NoStop}%
\bibitem [{\citenamefont {Ansel~et al.}(2024)}]{pytorch}%
  \BibitemOpen
  \bibfield  {author} {\bibinfo {author} {\bibfnamefont {J.}~\bibnamefont
  {Ansel~et al.}},\ }\bibfield  {title} {\bibinfo {title} {{PyTorch 2: Faster
  Machine Learning Through Dynamic Python Bytecode Transformation and Graph
  Compilation}},\ }in\ \href {https://doi.org/10.1145/3620665.3640366} {\emph
  {\bibinfo {booktitle} {29th ACM International Conference on Architectural
  Support for Programming Languages and Operating Systems, Volume 2 (ASPLOS
  '24)}}}\ (\bibinfo  {publisher} {ACM},\ \bibinfo {year} {2024})\BibitemShut
  {NoStop}%
\bibitem [{\citenamefont {Roth}\ \emph {et~al.}(2012)\citenamefont {Roth},
  \citenamefont {Mecke},\ and\ \citenamefont {Oettel}}]{martinsfunctional}%
  \BibitemOpen
  \bibfield  {author} {\bibinfo {author} {\bibfnamefont {R.}~\bibnamefont
  {Roth}}, \bibinfo {author} {\bibfnamefont {K.}~\bibnamefont {Mecke}},\ and\
  \bibinfo {author} {\bibfnamefont {M.}~\bibnamefont {Oettel}},\ }\bibfield
  {title} {\bibinfo {title} {{Communication: Fundamental measure theory for
  hard disks: Fluid and solid}},\ }\href {https://doi.org/10.1063/1.3687921}
  {\bibfield  {journal} {\bibinfo  {journal} {J. Chem. Phys.}\ }\textbf
  {\bibinfo {volume} {136}},\ \bibinfo {pages} {081101} (\bibinfo {year}
  {2012})}\BibitemShut {NoStop}%
\bibitem [{\citenamefont {Thorneywork}\ \emph {et~al.}(2014)\citenamefont
  {Thorneywork}, \citenamefont {Roth}, \citenamefont {Aarts},\ and\
  \citenamefont {Dullens}}]{thorneywork2014}%
  \BibitemOpen
  \bibfield  {author} {\bibinfo {author} {\bibfnamefont {A.~L.}\ \bibnamefont
  {Thorneywork}}, \bibinfo {author} {\bibfnamefont {R.}~\bibnamefont {Roth}},
  \bibinfo {author} {\bibfnamefont {D.~G. A.~L.}\ \bibnamefont {Aarts}},\ and\
  \bibinfo {author} {\bibfnamefont {R.~P.~A.}\ \bibnamefont {Dullens}},\
  }\bibfield  {title} {\bibinfo {title} {Communication: Radial distribution
  functions in a two-dimensional binary colloidal hard sphere system},\ }\href
  {https://doi.org/10.1063/1.4872365} {\bibfield  {journal} {\bibinfo
  {journal} {J. Chem. Phys.}\ }\textbf {\bibinfo {volume} {140}},\ \bibinfo
  {pages} {161106} (\bibinfo {year} {2014})}\BibitemShut {NoStop}%
\bibitem [{\citenamefont {Martin}\ \emph {et~al.}(2022)\citenamefont {Martin},
  \citenamefont {Hansen-Goos}, \citenamefont {Roth},\ and\ \citenamefont
  {Laird}}]{martin2022}%
  \BibitemOpen
  \bibfield  {author} {\bibinfo {author} {\bibfnamefont {S.~C.}\ \bibnamefont
  {Martin}}, \bibinfo {author} {\bibfnamefont {H.}~\bibnamefont {Hansen-Goos}},
  \bibinfo {author} {\bibfnamefont {R.}~\bibnamefont {Roth}},\ and\ \bibinfo
  {author} {\bibfnamefont {B.~B.}\ \bibnamefont {Laird}},\ }\bibfield  {title}
  {\bibinfo {title} {Inside and out: Surface thermodynamics from positive to
  negative curvature},\ }\href {https://doi.org/10.1063/5.0099295} {\bibfield
  {journal} {\bibinfo  {journal} {J. Chem. Phys.}\ }\textbf {\bibinfo {volume}
  {157}},\ \bibinfo {pages} {054702} (\bibinfo {year} {2022})}\BibitemShut
  {NoStop}%
\bibitem [{\citenamefont {Bernard}\ and\ \citenamefont
  {Krauth}(2011)}]{bernard2011}%
  \BibitemOpen
  \bibfield  {author} {\bibinfo {author} {\bibfnamefont {E.~P.}\ \bibnamefont
  {Bernard}}\ and\ \bibinfo {author} {\bibfnamefont {W.}~\bibnamefont
  {Krauth}},\ }\bibfield  {title} {\bibinfo {title} {Two-step melting in two
  dimensions: first-order liquid-hexatic transition},\ }\href
  {https://doi.org/10.1103/PhysRevLett.107.155704} {\bibfield  {journal}
  {\bibinfo  {journal} {Phys. Rev. Lett,}\ }\textbf {\bibinfo {volume} {107}},\
  \bibinfo {pages} {155704} (\bibinfo {year} {2011})}\BibitemShut {NoStop}%
\bibitem [{\citenamefont {Keskar}\ \emph {et~al.}(2017)\citenamefont {Keskar},
  \citenamefont {Mudigere}, \citenamefont {Nocedal}, \citenamefont
  {Smelyanskiy},\ and\ \citenamefont {Tang}}]{batchsize1}%
  \BibitemOpen
  \bibfield  {author} {\bibinfo {author} {\bibfnamefont {N.~S.}\ \bibnamefont
  {Keskar}}, \bibinfo {author} {\bibfnamefont {D.}~\bibnamefont {Mudigere}},
  \bibinfo {author} {\bibfnamefont {J.}~\bibnamefont {Nocedal}}, \bibinfo
  {author} {\bibfnamefont {M.}~\bibnamefont {Smelyanskiy}},\ and\ \bibinfo
  {author} {\bibfnamefont {P.~T.~P.}\ \bibnamefont {Tang}},\ }\href
  {https://arxiv.org/abs/1609.04836} {\bibinfo {title} {On large-batch training
  for deep learning: Generalization gap and sharp minima}} (\bibinfo {year}
  {2017}),\ \Eprint {https://arxiv.org/abs/1609.04836} {arXiv:1609.04836
  [cs.LG]} \BibitemShut {NoStop}%
\bibitem [{\citenamefont {Qian}\ and\ \citenamefont
  {Klabjan}(2020)}]{batchsize2}%
  \BibitemOpen
  \bibfield  {author} {\bibinfo {author} {\bibfnamefont {X.}~\bibnamefont
  {Qian}}\ and\ \bibinfo {author} {\bibfnamefont {D.}~\bibnamefont {Klabjan}},\
  }\href {https://arxiv.org/abs/2004.13146} {\bibinfo {title} {The impact of
  the mini-batch size on the variance of gradients in stochastic gradient
  descent}} (\bibinfo {year} {2020}),\ \Eprint
  {https://arxiv.org/abs/2004.13146} {arXiv:2004.13146 [math.OC]} \BibitemShut
  {NoStop}%
\bibitem [{\citenamefont {Lin}\ and\ \citenamefont {Oettel}(2018)}]{sam2018}%
  \BibitemOpen
  \bibfield  {author} {\bibinfo {author} {\bibfnamefont {S.-C.}\ \bibnamefont
  {Lin}}\ and\ \bibinfo {author} {\bibfnamefont {M.}~\bibnamefont {Oettel}},\
  }\bibfield  {title} {\bibinfo {title} {Phase diagrams and crystal-fluid
  surface tensions in additive and nonadditive two-dimensional binary hard-disk
  mixtures},\ }\href {https://doi.org/10.1103/PhysRevE.98.012608} {\bibfield
  {journal} {\bibinfo  {journal} {Phys. Rev. E}\ }\textbf {\bibinfo {volume}
  {98}},\ \bibinfo {pages} {012608} (\bibinfo {year} {2018})}\BibitemShut
  {NoStop}%
\end{thebibliography}%

\end{document}